\pdfoutput=1  
\def\nk{n_\mathrm{K}}
\def\acap{\\ \nonumber \\}

\def\rfr#1{Equation\,(\ref{#1})}
\def\rfrs#1#2{Equations\,(\ref{#1})--(\ref{#2})}

\def\Rfrs#1#2{Equations\,(\ref{#1})--(\ref{#2})}

\def\dert#1#2{\frac{{{\textrm{d}}}{#1}}{{{\textrm{d}}}{#2}}}
\def\virg#1{``#1"}
\def\eqi{\begin{equation}}
\def\eqf{\end{equation}}
\def\rp#1#2{\frac{#1}{#2}}
\def\lb#1{\label{#1}}
\def\bds#1{\boldsymbol{#1}}
\def\ton#1{\left(#1\right)}
\def\qua#1{\left[#1\right]}
\def\grf#1{\left\{#1\right\}}

\RequirePackage[2020-02-02]{latexrelease}
\documentclass{aastex631}
\usepackage{morefloats}
\usepackage[title]{appendix}
\usepackage{textcomp}
\usepackage{booktabs}
\usepackage{multirow}
\usepackage{rotating,tabularx}
\usepackage{float}
\usepackage{enumerate}
\usepackage{rotating}
\usepackage[polutonikogreek,english]{babel}
\usepackage{amsmath,starfont,textgreek,w-greek}
\usepackage[flushleft]{threeparttable}
\usepackage{amsthm}
\usepackage{bigints}
\usepackage{amscd}
\usepackage[mathlines]{lineno}
\usepackage{amssymb,dsfont}
\usepackage{graphicx,epsfig}
\usepackage{txfonts}
\usepackage{xr-hyper}
\usepackage{hyperref}
\usepackage[utf8]{inputenc}
\usepackage{newunicodechar,graphicx}

\usepackage[nointegrals]{wasysym}
\usepackage[caption=false]{subfig}

\allowdisplaybreaks

\makeatletter
\DeclareRobustCommand\ref{%
    \@ifstar\@refstar\T@ref
  }%
  \DeclareRobustCommand\pageref{%
    \@ifstar\@pagerefstar\T@pageref
  }%
 \makeatother

\begin{document}

\title{Nethotrons: exploring the possibility of measuring relativistic spin precessions\textcolor{black}{,} from Earth's satellites to the Galactic Centre}

\shortauthors{L. Iorio}

\author[0000-0003-4949-2694]{Lorenzo Iorio}
\affiliation{Ministero dell' Istruzione e del Merito. Viale Unit\`{a} di Italia 68, I-70125, Bari (BA),
Italy}

\email{lorenzo.iorio@libero.it}

\begin{abstract}
By \virg{nethotron}, from the ancient Greek verb for \virg{to spin}, it is meant here a natural or artificial rotating object, like a pulsar or an artificial satellite, whose rotational axis is cumulatively displaced by the post-Newtonian static (gravitoelectric) and stationary (gravitomagnetic) components of the gravitational field of some massive body around which it freely moves. Until now, both relativistic effects  have been measured only by the dedicated space-based mission Gravity Probe B in the terrestrial environment. It detected the gravitoelectric de Sitter and gravitomagnetic Pugh-Schiff spin precessions of four superconducting gyroscopes accumulated in a year after about 50 years from conception to  completion of data analysis at a cost of 750 million dollars to $0.3$ and $19$ per cent accuracy, respectively. The perspectives to measure them also with long-lived Earth's laser-ranged geodetic satellites, like those of the LAGEOS family or possibly one or more of them to be built specifically from scratch, and pulsars orbiting the supermassive black hole in the Galactic Centre, yet to be discovered, are preliminarily investigated. The double pulsar PSR J0737-3039A/B is examined as well.
\end{abstract}


%
%
%
\section{Introduction}

One of the most intriguing, and, at least to a certain extent, less tested  predictions of the General Theory of Relativity (GTR) is the temporal evolution of the spin axis $\bds{\hat{S}}$ of a gyroscope freely moving in the external gravitational field of a massive, also rotating body of mass $M$, equatorial radius $R$, quadrupole mass moment $J_2$ and angular momentum $\boldsymbol{J}$. Indeed, Newtonian mechanics implies that the spin of the orbiting object experiences a steady, cumulative gravitational precession with respect to distant observers, induced solely by $M$, only if it is itself nonspherical being endowed with, e.g., an own oblateness $J_2^\mathrm{s}$ \cite{1975PhRvD..12..329B,2011CeMDA.111..105C}. A further Newtonian contribution of gravitational origin to the overall gyroscope's spin precession may arise if a distant third body is present as well \cite{2011CeMDA.111..105C}. Instead, according to GTR, $\bds{\hat{S}}$ changes cumulatively its orientation with respect to remote stars even if the particle carrying it is spherically symmetric and if there are no other bodies to pull it other than its primary. Furthermore, both $M$ and $\boldsymbol{J}$ contribute to the total gyroscope's spin rate through the static and stationary post-Newtonian (pN) components of the exterior gravitational field of the central source of which they are the mass and angular momentum, respectively. To the first post-Newtonian (1pN) level,
the gravitoelectric and gravitomagnetic precessions of the spin of a pointlike gyroscope of negligible mass due to $M$ and $\boldsymbol{J}$ were calculated by de Sitter \cite{1916MNRAS..77..155D} and Fokker \cite{1921KNAB...23..729F}, and Pugh and Schiff \cite{Pugh59,Schiff60}, respectively. They can be found written in compact, vectorial form, e.g., in\footnote{Such papers contain expressions which are also valid in the case of a two-body system whose members have both comparable masses, angular momenta and quadrupole moments; see also \cite{DamRu74}.} \cite{1975PhRvD..12..329B,1979GReGr..11..149B}.

At present, the only direct test of both the aforementioned pN gravitoelectromagnetic effects in a dedicated, controlled experiment was conducted in the Earth's surrounding by the Gravity Probe B (GP-B) space-based mission \cite{2015CQGra..32v4001E}. The de Sitter (dS) and Pugh-Schiff (PS) precessions of four superconducting spinning spheres carried onboard a drag-free spacecraft orbiting our planet in a low-altitude, polar orbit were successfully measured to a $0.3$ and $19$ per cent accuracy \cite{2011PhRvL.106v1101E}, respectively, compared to an initially estimated much better one \cite{Varenna74,2001LNP...562...52E}. The entire mission, whose duration since the initial conception to completion of data analysis can be estimated to have been around 50 years, cost about 750 million dollars; the data collection phase lasted a year \cite{2011PhyOJ...4...43W}. At present, the GP-B remains the only test of the PS spin precession and the only undisputed test ever of a gravitomagnetic effect. As far as the dS precession is concerned, it was measured also with the Lunar Laser Ranging (LLR) technique \cite{2021Univ....7...34B} by considering the whole orbit of the Earth-Moon system as a giant gyroscope whose orientation with respect to distant stars is cumulatively displaced during its freely falling in the deformed spacetime of the Sun; the latest tests are accurate to $0.4$ \cite{2009IAU...261.0801W} and $0.09$ per cent \cite{2018CQGra..35c5015H}. The generalization of the dS spin precession to a two-body system of comparable masses was measured to a $\simeq 13-6$ per cent accuracy \cite{2008Sci...321..104B,2024A&A...682A..26L} with the double pulsar PSR J0737–3039A/B \cite{2003Natur.426..531B,2004Sci...303.1153L} whose members can be viewed, among other things, as strongly self-gravitating gyroscopes, contrary to the man-made light spheres of GP-B  \cite{2024LRR....27....5F}.

\textcolor{black}{The s}cope of this paper is to explore the possibility of performing further tests of the 1pN spin precessions with either artificial or natural pointlike gyroscopes by exploiting existing objects, possibly still to be discovered, or by specifically building new ones that are not excessively expensive. It should be recalled that, in view of its complexity and cost, GP-B is most likely never to be repeated.
The role of the negligible mass gyroscope orbiting a massive, rotating primary will be played by both Earth's artificial satellites and pulsars orbiting a supermassive black hole (SMBH), collectively denoted as \virg{nethotrons}, from the ancient Greek verb $\upnu\acute{\upeta}\uptheta\upomega$ (\textit{n$\acute{\bar{e}}$th$\bar{o}$}), meaning \virg{to spin}, and the suffix -$\uptau\uprho\upo\upnu$ (-\textit{tron}) used to make instrument nouns.

As for artificial nethotrons, the first thought immediately goes to geodetic satellites of the LAGEOS type \citep{2019JGeod..93.2181P} tracked with the Satellite Laser Ranging (SLR) technique \citep{SLR11}, the evolution of whose spin axes has actually been measured for a long time \cite{2000ITGRS..38.1417O,2001AdSpR..28.1309V,2004JGRB..109.6403A,2004ITGRS..42..202O,2007ITGRS..45..201K,2009AdSpR..44..621K,2010AdSpR..46..251K,2011AdSpR..48..343K,2012AdSpR..50.1473K,2013AdSpR..52.1332K,2013AdSpR..51..162K,2014AdSpR..54.2309K,2016AdSpR..57..983K,2017AdSpR..60.1389K,2024AdSpR..74.5725K}.
Given the relatively modest accuracy of the test of the gravitomagnetic spin precession performed with the very sophisticated and expensive GP-B, built expressly for that purpose, even just the idea of using the simple passive geodetic satellites might seem decidedly bizarre, if not even absurd. Nevertheless, it is  anyway explored here for the following reasons: a) There are so many SLR satellites currently in orbit. b) They are expected to remain there for a virtually indefinite period of time. c) They are tracked on a nearly continuous basis. d) Measuring their spin motion is currently one of the main goals of satellite geodesy. e) The temporal signatures of the pN effects of interest, being generally different from linear trends in view of the non-dedicated satellites' orbital and spin configurations, may make their detection easier, allowing also for an effective disentangling from the competing signals of classical origin. f) As if that were not enough, building and launching one or more new LAGEOS-type satellites is feasible at a relatively low cost. The latter would certainly not increase much even if particular care were taken in the manufacturing of such objects. g) Finally, it is anyway of interest to understand quantitatively where do we stand evaluating how much the measurement techniques of the SLR satellites' spin motion would have to become more precise to allow the measurement of the relativistic signals of interest in possibly implementing one or more low-cost versions of GP-B.

Instead, the best candidate for a natural nethotron would be a pulsar orbiting the SMBH in Sgr A$^\ast$ at the Galactic Centre (GC) in, say, a $0.5$ yr or a $2-5$ yr orbit, yet to be discovered; efforts towards this goal have been underway for a long time \cite{2014ApJ...784..106Z,2021ApJ...914...30L,2023PhRvD.108l4027C,2023ApJ...959...14T}. This eventuality has been long-awaited since it would allow to probe the spacetime\footnote{In addition to GTR features, pulsars around Sgr A$^\ast$ may be used to test also modified models of gravity \cite{2022JCAP...11..051D,2024JCAP...04..087H} and dark matter distribution in the GC \cite{2023PhRvD.108l3034H}.} around the hole, with particular attention to its angular momentum $\boldsymbol{J}$ \cite{1999ApJ...514..388W,2004ApJ...615..253P,2012ApJ...747....1L,2016ApJ...818..121P,2017ApJ...849...33Z,2024PhRvL.133w1402H}. To this aim, efforts have been focused so far on the secular orbital precessions of the pulsar.
Measuring also the 1pN pulsar's spin precession(s) would provide, in principle, a further tool to effectively constrain the key parameters of the SMBH's spacetime which, in the case of a rotating hole \cite{1970Natur.226...64B}, is believed to be described by the Kerr metric \cite{1963PhRvL..11..237K,2015CQGra..32l4006T}. In principle, the internal structure of a neutron star couples with the external deformed spacetime impacting on both its orbital and rotational dynamics \cite{2006PhRvD..74l4006M,2014PhRvD..90f4035H,2016PhRvD..94d4008R}. Nonetheless, in the present case, the nethotron approximation is valid also for it since it will be shown that the multipolar structure of the pulsar's self-gravity is negligible for its spin precession already at the Newtonian level.

The paper is organized as follows. The analytical expressions of the 1pN dS and PS spin precessions  along with those of the Newtonian one due to the nethotron's own oblateness are provided in Section\,\ref{Sec:1}. The pN spin axis shifts of the existing SLR satellites LAGEOS, LAGEOS 2 and LARES are numerically produced and displayed in Section\,\ref{Sec:2}. In it, also the impact of the satellite's own oblateness on the evolution of its spin is treated.  Moreover, the pN precessions of a hypothetical new satellite, dubbed NethoSAT (NS), are obtained as well. Section\,\ref{Sec:3} deals with the case of a pulsar orbiting the SMBH in Sgr A$^\ast$. The case of the double pulsar is  treated in Section\,\ref{Sec:4}. Section\,\ref{Sec:5} summarizes the findings and offers conclusions.
\section{Analytical expressions of the Newtonian and post-Newtonian spin precessions}\label{Sec:1}

In the following, $G$ is the Newtonian constant of universal gravitation, $c$ is the speed of light in vacuum, $\upmu:=GM$ is the standard gravitational parameter of the primary whose mass is $M$, $a$ is the semimajor axis, $\nk:=\sqrt{\upmu/a^3}$ is the Keplerian mean motion,  $e$ is the eccentricity, $p:=a\ton{1-e^2}$ is the semilatus rectum, $I$ is the inclination of the orbital plane to the reference plane, and $\Omega$ is the longitude of the ascending node. Furthermore, the orbital unit vectors
\begin{align}
\bds{\hat{l}} & = \grf{\cos\Omega,\sin\Omega,0},\acap
\bds{\hat{m}} & = \grf{-\cos I\sin\Omega,\cos I\cos\Omega,\sin I},\acap
\bds{\hat{h}} & = \grf{\sin I\sin\Omega,-\sin I\cos\Omega,\cos I}
\end{align}
are defined in such a way that $\bds{\hat{l}}\,\bds\times\,\bds{\hat{m}} = \bds{\hat{h}}$. While $\bds{\hat{h}}$ is aligned with the orbital angular momentum, $\bds{\hat{l}}$ and $\bds{\hat{m}}$ lie within the orbital plane, being $\bds{\hat{l}}$ directed along the line\footnote{It is the intersection of the orbital plane with the fundamental one.} of nodes toward the ascending node.

By parameterizing the nethotron's spin axis $\bds{\hat{S}}$ in terms of its right ascension (RA) $\alpha$ and declination (decl.) $\delta$, it can be written as
\begin{align}
\hat{S}_x &= \cos\alpha\cos\delta, \acap
\hat{S}_y &= \sin\alpha\cos\delta, \acap
\hat{S}_z &= \sin\delta.
\end{align}
Thus, the rates of change of $\alpha$ and $\delta$ can be expressed as
\begin{align}
\dert\alpha t \lb{dRAdt} &= \textcolor{black}{
\rp{\cos^2\alpha}{{\hat{S}_x}^2}\ton{\dert{\hat{S}_y}{t}\hat{S}_x - \dert{\hat{S}_x}{t}\hat{S}_y
}}, \\
\dert\delta t \lb{ddecdt} &= \rp{1}{\cos\delta}\dert{\hat{S}_z}{t}.
\end{align}
In view of the following developments, it is useful to introduce the angle
\eqi
\eta:=\alpha-\Omega\lb{etaaO},
\eqf
which is the longitude of the ascending node reckoned from the projection of the spin axis onto the reference $\Pi$ plane.

It should be stressed that, in any realistic scenario, the orientation of the nethotron's orbital plane, characterized by $I$ and $\Omega$, does generally change after every orbit because of a variety of physical effects, of gravitational and non-gravitational origin, the most important of which is the quadrupole mass moment $J_2$ of the primary. Next, one finds the gravitomagnetic field itself  and possibly the pull of a distant third body.
The first two perturbations give rise to \cite{2024gpno.book.....I}
\begin{align}
\dert I t \lb{dIdt}& = -\rp{3}{2}\nk J_2\ton{\rp{R}{p}}^2\texttt{Jh}\texttt{Jl} + \rp{2GJ\texttt{Jl}}{c^2 a^3\ton{1 - e^2}^{3/2}}, \acap
\dert\Omega t \lb{dOdt}& = -\rp{3}{2}\nk J_2\csc I\ton{\rp{R}{p}}^2\texttt{Jh}\texttt{Jm} + \rp{2GJ\csc I \texttt{Jm}}{c^2 a^3\ton{1 - e^2}^{3/2}},
\end{align}
where
\begin{align}
\mathtt{Jl} \lb{Jm} &:= \bds{\hat{J}}\boldsymbol\cdot\bds{\hat{l}}, \acap
\mathtt{Jm} \lb{Jl} &:= \bds{\hat{J}}\boldsymbol\cdot\bds{\hat{m}}, \acap
\mathtt{Jh} \lb{Jh} &:= \bds{\hat{J}}\boldsymbol\cdot\bds{\hat{h}}.
\end{align}
\Rfrs{dIdt}{dOdt} hold for any orbital configuration and for a generic orientation of the primary's spin axis in space.
Since $I$ and $\Omega$ enter explicitly \rfrs{dRAdt}{ddecdt}, which are mutually coupled, one containing the variable of the other, the signatures of  $\alpha\ton{t}$ and $\delta\ton{t}$ are generally not secular trends, being instead long-term harmonic  time series modulated by the characteristic time scales of the averaged variations of the orbital plane.

\subsection{The gravitoelectric de Sitter spin precession}

The averaged gravitoelectric dS spin rate is \cite{1975PhRvD..12..329B,1979GReGr..11..149B}
\eqi
\dert{\bds{\hat{S}}}{t} = {\boldsymbol{\Omega}}_\mathrm{dS}\,\bds\times\,\bds{\hat{S}}.\lb{Sdot_dS}
\eqf
In it, the precession velocity vector is given by
\eqi
{\boldsymbol{\Omega}}_\mathrm{dS} = \mathcal{A}_\mathrm{dS}\,\bds{\hat{h}},\lb{OdS}
\eqf
where
\eqi
\mathcal{A}_\mathrm{dS} = \rp{3\nk\upmu}{2c^2 p}.\lb{AdS}
\eqf
The resulting spin's RA and decl. averaged precessions per orbit are
\begin{align}
\dert\alpha t \lb{RAdS}& = \mathcal{A}_\mathrm{dS}\ton{\cos I + \sin I \sin\eta\tan\delta}, \\
\dert\delta t \lb{DECdS}& = \mathcal{A}_\mathrm{dS}\sin I\cos\eta.
\end{align}

The dS spin precession depends only on the nethotron's own spin-orbit coupling; it is zero when the orbital and spin angular momenta of the satellite are parallel, being, instead, maximum when they are mutually perpendicular.

\subsection{The gravitomagnetic Pugh-Schiff spin precession}

The gravitomagnetic PS spin precession is \cite{1975PhRvD..12..329B,1979GReGr..11..149B}
\eqi
\dert{\bds{\hat{S}}}{t} = {\boldsymbol{\Omega}}_\mathrm{PS}\,\bds\times\,\bds{\hat{S}}.\lb{Sdot_PS}
\eqf
In it, the precession velocity vector is given by
\eqi
{\boldsymbol{\Omega}}_\mathrm{PS} = \rp{\mathcal{A}_\mathrm{PS}}{2}\qua{3\ton{\mathtt{Jl}\,\bds{\hat{l}} + \mathtt{Jm}\,\bds{\hat{m}}} - 2\bds{\hat{J}}},\lb{OPS}
\eqf
where $\bds{\hat{J}}$ is the primary's spin axis, parameterized in terms of its RA $\alpha_J$ and decl. $\delta_J$ as
\begin{align}
\hat{J}_x &= \cos\alpha_J\cos\delta_J, \acap
\hat{J}_y &= \sin\alpha_J\cos\delta_J, \acap
\hat{J}_z &= \sin\delta_J,
\end{align}
and
\eqi
\mathcal{A}_\mathrm{PS} \lb{APS} := \rp{GJ}{c^2 a^3\ton{1-e^2}^{3/2}}.
\eqf
The resulting spin's RA and decl. averaged precessions per orbit are
\begin{align}
\dert\alpha t \nonumber  &= -\rp{\mathcal{A}_\mathrm{PS}}{4}\ton{\cos\delta_J\grf{-3 \sin 2I \sin\eta_J  + \qua{2 \cos\eta  \cos\eta_J   + \ton{-1 + 3 \cos 2I} \sin\eta \sin\eta_J } \tan\delta } + \right.\\
\lb{RALT}&\left. + \sin\delta_J  \qua{1 - 3 \sin^2 I + 3 \cos I \ton{\cos I + 2 \sin I \sin\eta \tan\delta}}}, \\
\dert\delta t \lb{DECLT} & = \rp{\mathcal{A}_\mathrm{PS}}{4}\grf{2 \cos\delta_J  \cos\eta_J  \sin\eta +
 \cos\eta \qua{-3 \sin 2I \sin\delta_J  + \ton{1 - 3 \cos 2I} \cos\delta_J  \sin\eta_J }},
\end{align}
where
\eqi
\eta_J:=\alpha_J - \Omega,
\eqf
analogously to \rfr{etaaO}.
If, as in the case of the Earth, $\delta_J=90^\circ$, \rfrs{RALT}{DECLT} reduce to
\begin{align}
\dert\alpha t \lb{RALT2} &= -\rp{\mathcal{A}_\mathrm{PS}}{4}\qua{1 - 3 \sin^2 I + 3 \cos I \ton{\cos I + 2 \sin I \sin\eta \tan\delta}}, \\
\dert\delta t \lb{DECLT2} & = -\rp{3\mathcal{A}_\mathrm{PS}}{4}\sin 2I\cos\eta.
\end{align}

\subsection{The Newtonian spin precession due to the gyroscope's own oblateness}

The Newtonian gravitational spin precession induced by the coupling of primary's mass $M$ with the satellite's quadrupole mass moment $J_2^\mathrm{s}$ is \cite{1975PhRvD..12..329B,1979GReGr..11..149B}
\eqi
\dert{\bds{\hat{S}}}{t} = {\boldsymbol{\Omega}}_{J_2^\mathrm{s}}\,\bds\times\,\bds{\hat{S}}.\lb{Sdot_Qs}
\eqf
In it, the precession velocity vector is given by
\eqi
{\boldsymbol{\Omega}}_{J_2^\mathrm{s}} = \mathcal{A}_{J_2^\mathrm{s}}\ton{\bds{\hat{S}} - 3\,\mathtt{Sh}\,\bds{\hat{h}}}, \lb{O_Qs}
\eqf
where
\begin{align}
\mathcal{A}_{J_2^\mathrm{s}} \lb{AJ2} & := \rp{5\upmu J_2^\mathrm{s}}{4\omega_\mathrm{s} a^3\ton{1-e^2}^{3/2}},\\
\mathtt{Sh} & := \bds{\hat{h}}\bds\cdot\bds{\hat{S}}.
\end{align}
In \rfr{AJ2}, the expression valid for a homogenous sphere was assumed for the nethotron's moment of inertia, and
\eqi
\omega_\mathrm{a} = \rp{2\pi}{P_\mathrm{s}},
\eqf
where $P_\mathrm{s}$ is its spin period, assumed constant over an orbital revolution.
From \rfrs{Sdot_Qs}{O_Qs}, it turns out that the classical spin rate due to the nethotron's own oblateness vanishes if its spin axis is either parallel or perpendicular to its orbital angular momentum.

The resulting spin's RA and decl. averaged precessions per orbit are
\begin{align}
\dert\alpha t \lb{RAJ2} & = \rp{3}{4}\mathcal{A}_{J_2^\mathrm{s}}\qua{-\ton{1 + 3 \cos 2 I + 2\sin^2 I\cos 2\eta}\sin\delta
+ 2\sin 2 I \sin\eta\cos 2\delta\sec\delta}, \\
\dert\delta t \lb{DECJ2} & = -\rp{3}{2}\mathcal{A}_{J_2^\mathrm{s}}\cos\eta\ton{\sin 2 I \sin\delta - 2\sin^2 I \sin\eta\cos\delta}.
\end{align}

%

\section{The case of the SLR satellites}\lb{Sec:2}

From \rfrs{Sdot_dS}{OdS}, \rfrs{Sdot_PS}{OPS} and \rfrs{Sdot_Qs}{O_Qs} it turns out that if the three angular momenta $\boldsymbol{J}, \bds h, \bds S$ are mutually aligned, as for a \textcolor{black}{nethotron} moving along an equatorial orbit with its spin axis $\boldsymbol{\hat{S}}$ parallel to the Earth's axis, all the precessions of Section \ref{Sec:1} vanish. If, instead, $\boldsymbol{\hat{S}}$ lies somewhere in the orbital plane parallel to the Earth's equatorial one, its RA experiences a combined pN secular shift dominated by the larger dS one, which is prograde, with respect to the smaller retrograde PS one. Both the pN rates depend only on the inclination of the orbital plane.
Note that, in this case, the orbital plane would stay parallel to the Earth's equator since no gravitational torques displace it, as shown by by \rfrs{dIdt}{Jh}. Furthermore, the nethotron's spin would remain perpendicular to the orbital angular momentum since it would move within the orbital plane, being only its RA affected by the overall pN effect. Thus, the Newtonian precession due to $J_2^\mathrm{a}$ would keep equal to zero. This fact would occur even if the pN effects did not exist at all. Indeed, $\delta$ is not classically shifted at all since, for $I=0^\circ$, the right-hand-side of \rfr{DECJ2} is identically zero over all the time. Thus, if the decl. of the nethotron's spin axis starts from $\delta_0=0^\circ$, it will remain zero forever. This implies that also the right-hand-side of \rfr{RAJ2} will be always zero by keeping $\alpha$ equal to its initial value $\alpha_0$ whatever it is.

Another orbital configuration which would ideally assure a constant orbital plane would be the polar one, i.e., when the Earth's spin axis lies within the orbital plane, as it can straightforwardly be inferred by \rfrs{dIdt}{Jh}. In this case, $\bds h$ and $\bds J$ are mutually perpendicular, and the largely dominant Newtonian precessions due to the primary's oblateness vanish. If $\boldsymbol{\hat{S}}$ starts aligned with $\boldsymbol{\hat{h}}$, which occurs for $\alpha_0=\Omega_0\pm 90^\circ,\,\delta_0=0^\circ$, it is slowly displaced parallel to the equatorial plane by the PS effect changing only its RA in the prograde sense.
If, instead, $\boldsymbol{\hat{S}}$ starts within the equatorial plane perpendicular to both $\boldsymbol{\hat{J}}$ and $\boldsymbol{\hat{h}}$, i.e. for $\alpha_0=\Omega_0\pm 180^\circ,\,\delta_0=0^\circ$, its RA undergoes the PS shift, while its decl. experiences the dS variation, as in the GP-B experiment.
However, in the case of a polar orbit, the Newtonian precession due to the nethotron's own oblateness would generally not vanish even if the pN effects were absent. Indeed, contrary to the equatorial case, the right-hand-sides of \rfrs{RAJ2}{DECJ2} are not identically zero.

The spin and orbital configurations of the existing SLR satellites of the LAGEOS-type, listed in Table \ref{Tab:1}, do not fall in any of the previous cases.
\begin{table}[h]
\caption{In the first three rows, the relevant spin and orbital parameters of LAGEOS, LAGEOS 2 and LARES are shown as per Table\,IV and Table\,V of \cite{2018PhRvD..98d4034V}. The values of the spin's periods $P_\mathrm{s}$, RA $\alpha$ and decl. $\delta$ refer to 15 May 1976 for LAGEOS, 23 October 1992 for LAGEOS 2 and 13 February 2012 for LARES. The figures for the rates of change $\dot P_\mathrm{s}$ of the satellites' spin periods are inferred from Figure 1, Figure 3 and Figure 5 of \cite{2018PhRvD..98d4034V}. The dimensionless quadrupole mass moments $J_2^\mathrm{s}$ for each satellite were obtained from the values for their masses $M_\mathrm{s}$, radii $R_\mathrm{s}$ and principal moments of inertia $\mathcal{I}^\mathrm{s}_{xx},\,\mathcal{I}^\mathrm{s}_{yy},\,\mathcal{I}^\mathrm{s}_{zz}$, reported in Table\,I of \cite{2018PhRvD..98d4034V} and in the body of that article, as $J_2^\mathrm{s} = \ton{2\mathcal{I}^\mathrm{ss}_{zz} - \mathcal{I}^\mathrm{s}_{xx} - \mathcal{I}^\mathrm{s}_{yy}}/2M_\mathrm{s}R_\mathrm{s}^2$ \cite{2005som..book.....C}.
The last row displays the orbital and spin parameters  of GP-B retrieved from \cite{Kahn07}.
}\lb{Tab:1}
\begin{center}
\begin{tabular}{|l|l|l|l|l|l|l|l|l|l|l|}
  \hline
Satellite & Semimajor axis $a$ & Eccentricity $e$ & Inclination $I$ & Node $\Omega$ & RA $\alpha$ & decl. $\delta$ &$P_\mathrm{s}$ & $\dot P_\mathrm{s}$ & $R_\mathrm{s}$ & $J_2^\mathrm{s} $\\
\hline
LAGEOS  & $12\,270$ km & $0.004$ & $109.84^\circ$ & $313.72^\circ$ & $150^\circ$ & $-68^\circ$ & $0.48$ s & $1.4\times 10^{-6}$ & $0.3$ m & $0.01256$\\
LAGEOS 2 & $12\,162$ km & $0.014$ & $52.66^\circ$ & $60.62^\circ$ & $230^\circ$ & $-81.8^\circ$ & $0.81$ s & $1.9\times 10^{-6}$ & $0.3$ m & $0.01234$\\
LARES & $7820.35$ km & $0.001$ & $69.49^\circ$ & $236.4^\circ$ & $186.5^\circ$ & $-73^\circ$ & $11.8$ s & $3\times 10^{-6}$ & $0.182$ m & $0.00078$\\
GP-B & $7027.4$ km & $0.0014$ & $90.007^\circ$ & $163.26^\circ$ & $\Omega + 180^\circ$ & $0^\circ$  & - & - & - & -\\
\hline
\end{tabular}
\end{center}
\end{table}

\begin{figure}[ht!]
\centering
\begin{tabular}{cC}
\includegraphics[width = 8.5 cm]{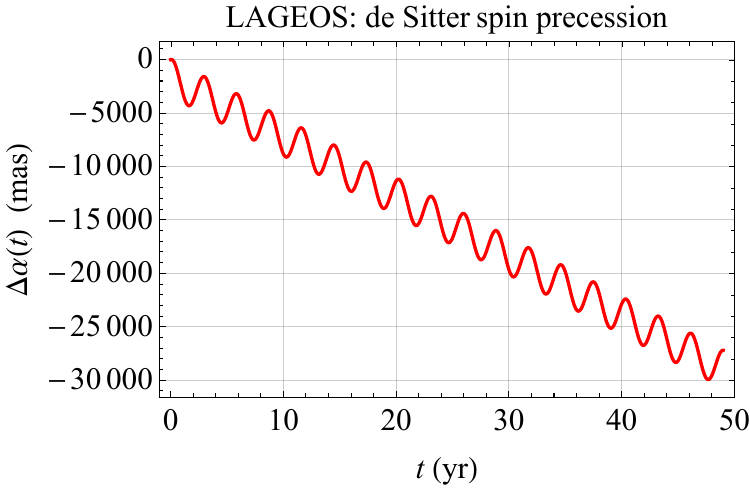} & \includegraphics[width = 8.5 cm]{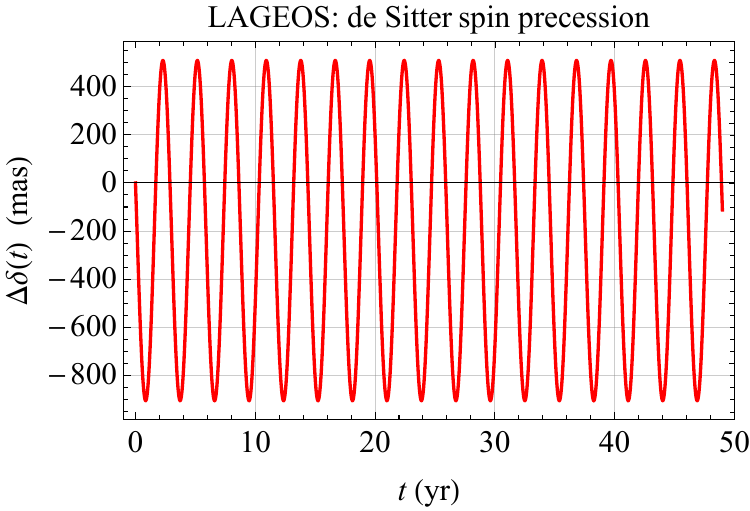}\\
\includegraphics[width = 8.5 cm]{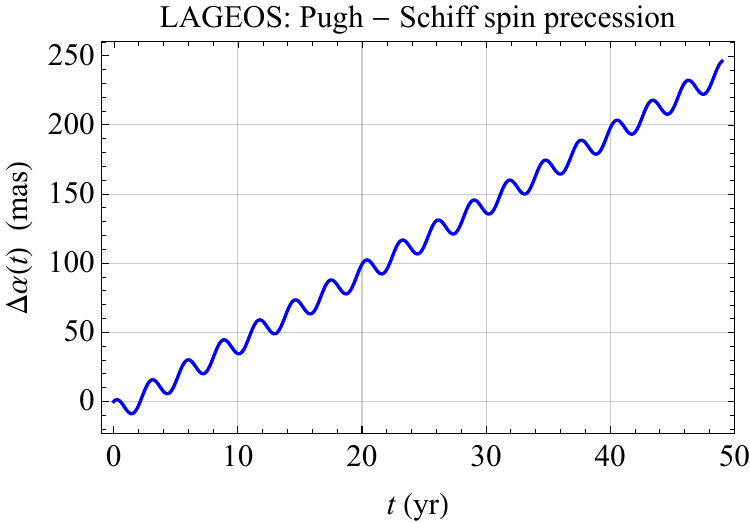} & \includegraphics[width = 8.5 cm]{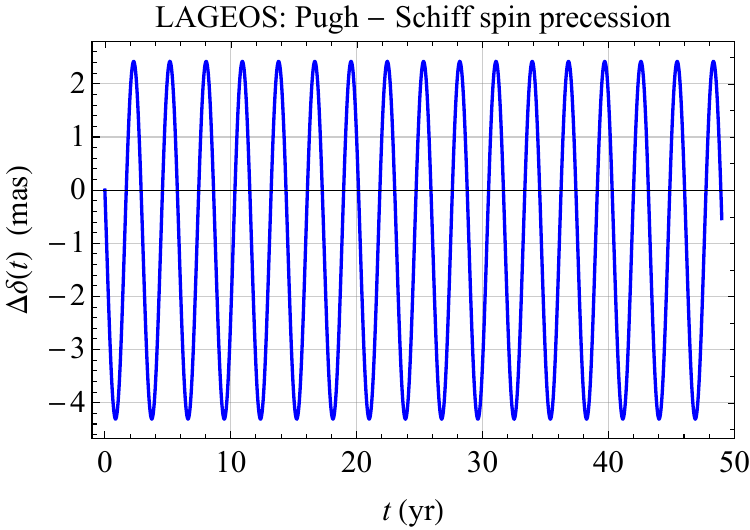}\\
\includegraphics[width = 8.5 cm]{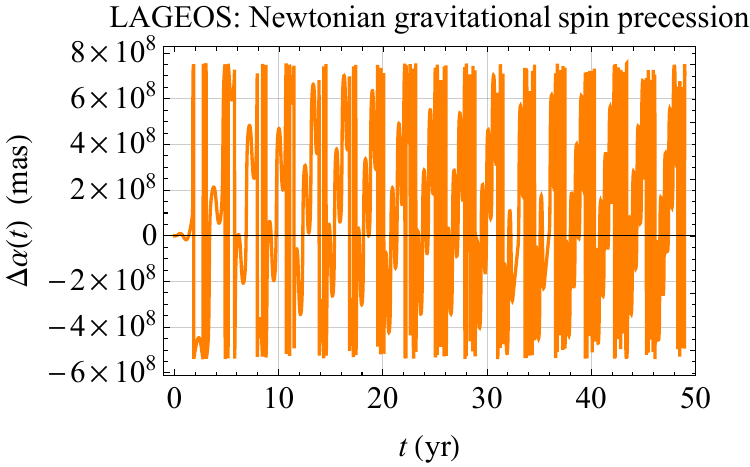} & \includegraphics[width = 8.5 cm]{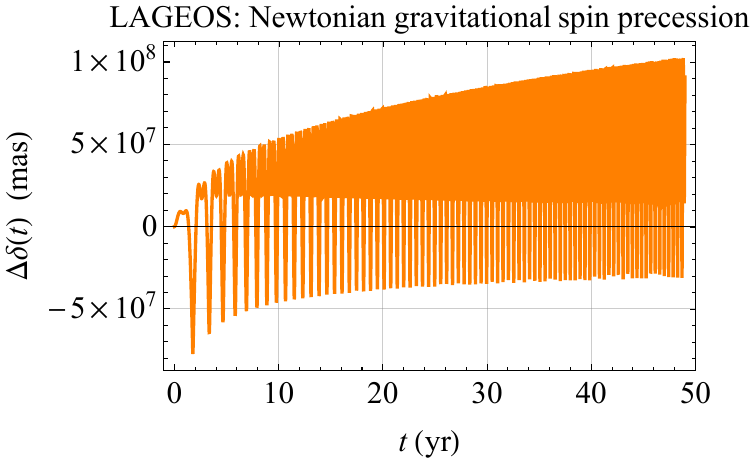}\\
\end{tabular}
\caption{
Numerically integrated time series, in mas, of the dS (upper row), PS (middle row) and Newtonian (lower row)  RA and decl. shifts $\Delta\alpha\ton{t}$ and $\Delta\delta\ton{t}$ of the spin axis of LAGEOS over 49 yr. The initial values of the spin and orbital parameters were retrieved from Table\,\ref{Tab:1}. For the satellite's spin period $P_\mathrm{s}$, entering the Newtonian shifts, a linearly varying model \cite{2018PhRvD..98d4034V} according to the values of Table\,\ref{Tab:1} for $\dot P_\mathrm{s}$ was adopted.
}\label{Fig:1}
\end{figure}

\begin{figure}[ht!]
\centering
\begin{tabular}{cC}
\includegraphics[width = 8.5 cm]{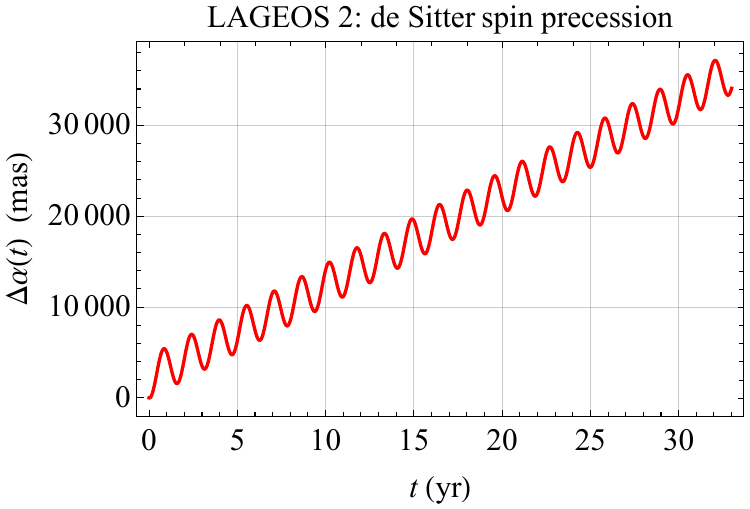} & \includegraphics[width = 8.5 cm]{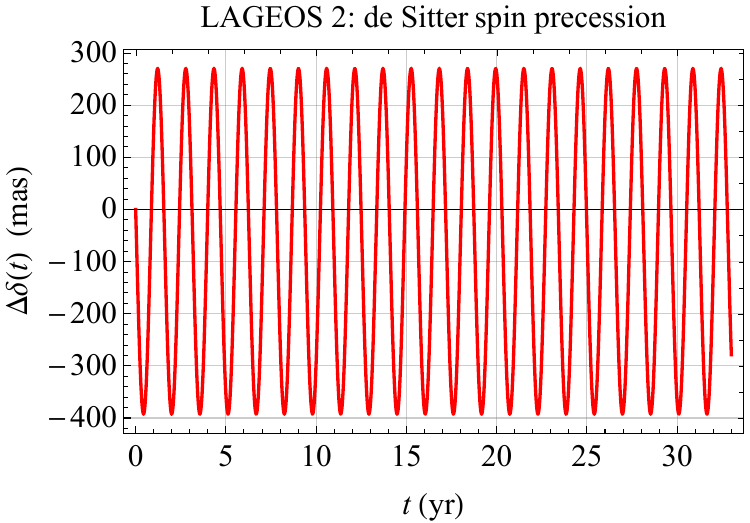}\\
\includegraphics[width = 8.5 cm]{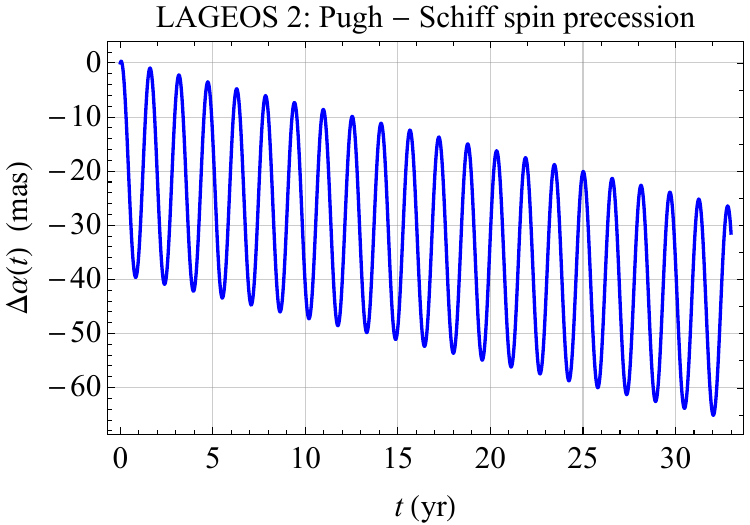} & \includegraphics[width = 8.5 cm]{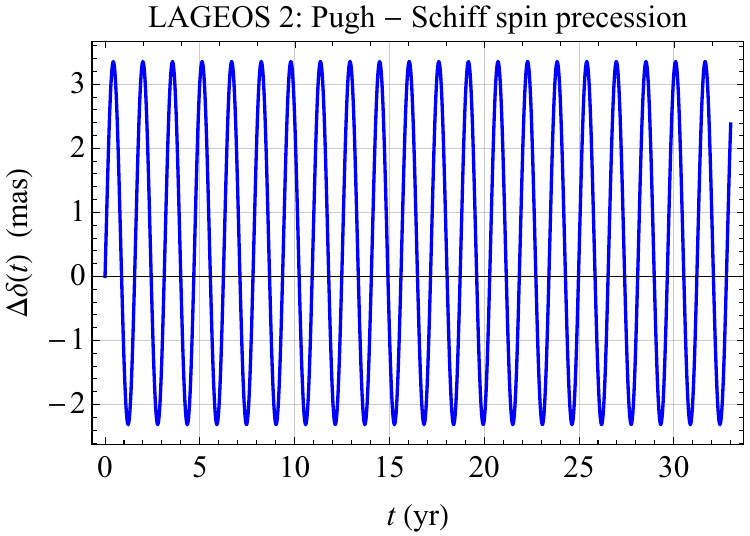}\\
\includegraphics[width = 8.5 cm]{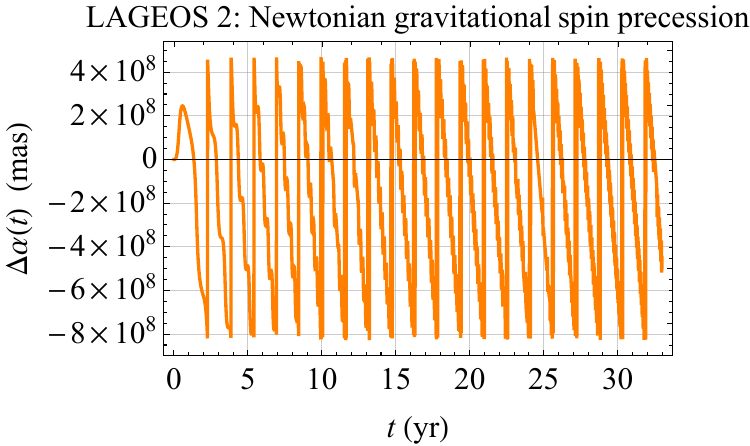} & \includegraphics[width = 8.5 cm]{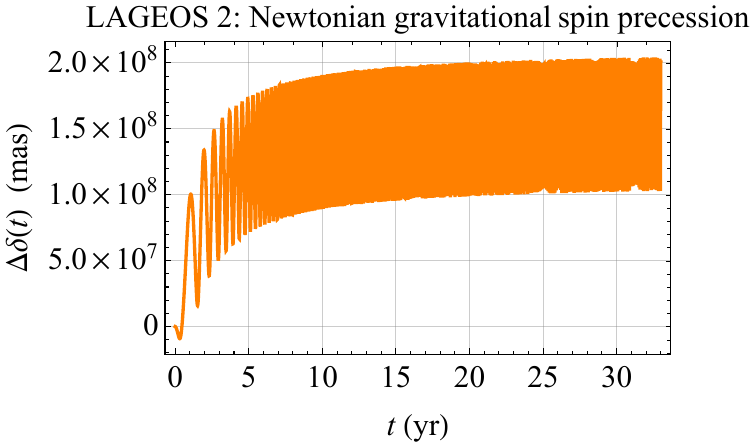}\\
\end{tabular}
\caption{
Numerically integrated time series, in mas, of the dS (upper row), PS (middle row) and Newtonian (lower row)  RA and decl. shifts $\Delta\alpha\ton{t}$ and $\Delta\delta\ton{t}$ of the spin axis of LAGEOS 2 over 33 yr. The initial values of the spin and orbital parameters were retrieved from Table\,\ref{Tab:1}. For the satellite's spin period $P_\mathrm{s}$, entering the Newtonian shifts, a linearly varying model \cite{2018PhRvD..98d4034V} according to the values of Table\,\ref{Tab:1} for $\dot P_\mathrm{s}$ was adopted.
}\label{Fig:2}
\end{figure}

\begin{figure}[ht!]
\centering
\begin{tabular}{cC}
\includegraphics[width = 8.5 cm]{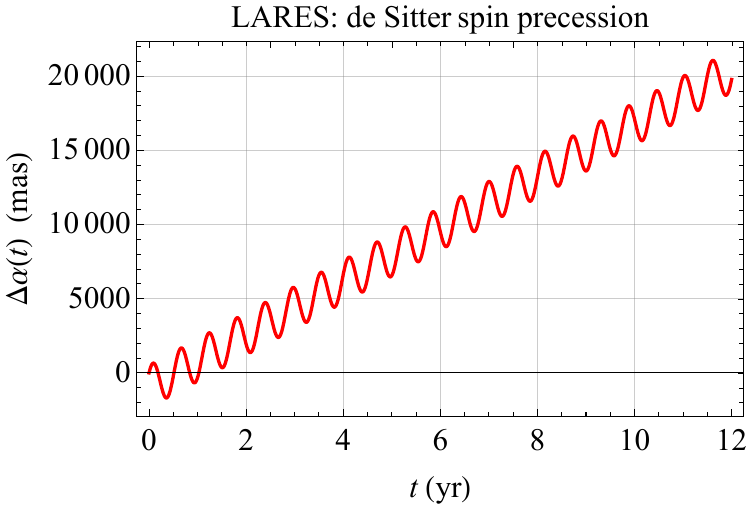} & \includegraphics[width = 8.5 cm]{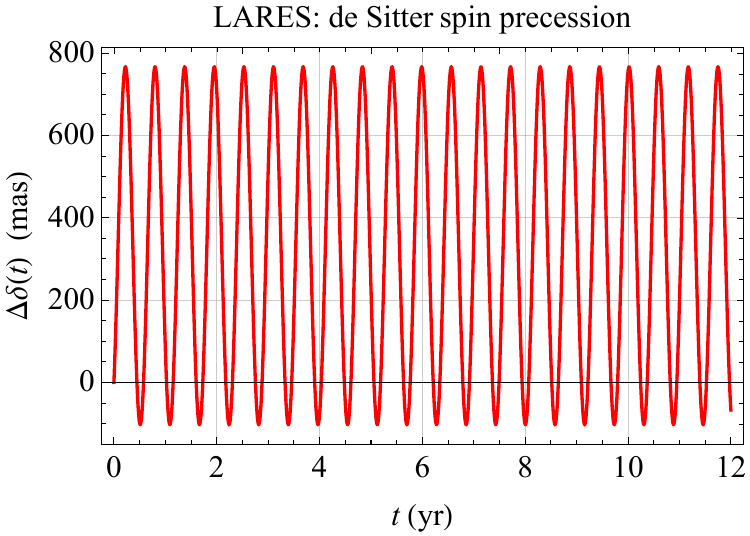}\\
\includegraphics[width = 8.5 cm]{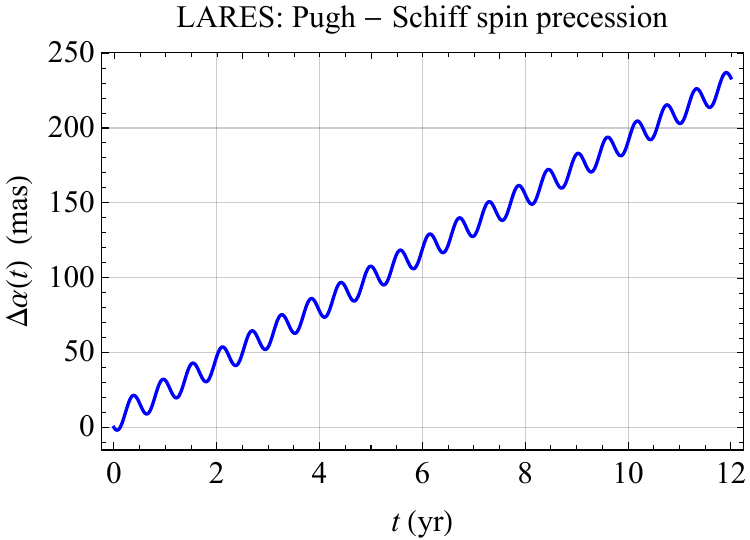} & \includegraphics[width = 8.5 cm]{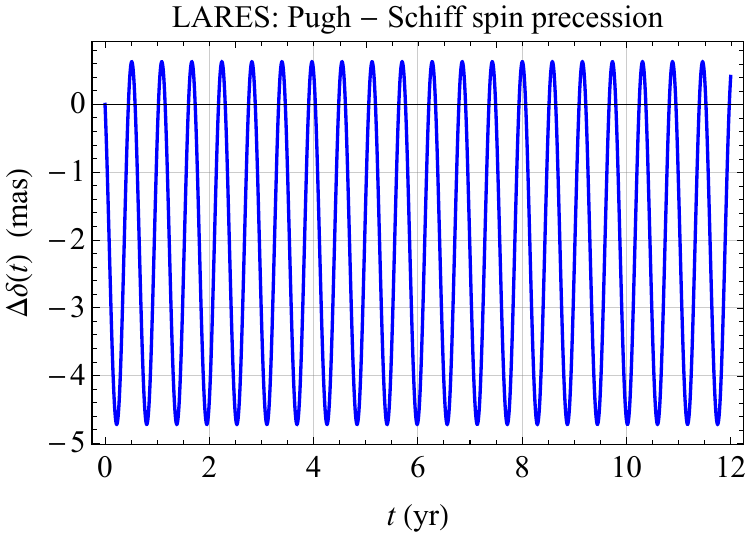}\\
\includegraphics[width = 8.5 cm]{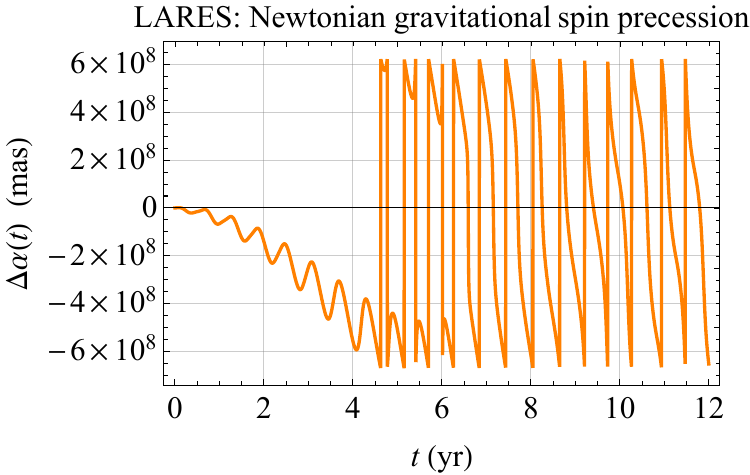} & \includegraphics[width = 8.5 cm]{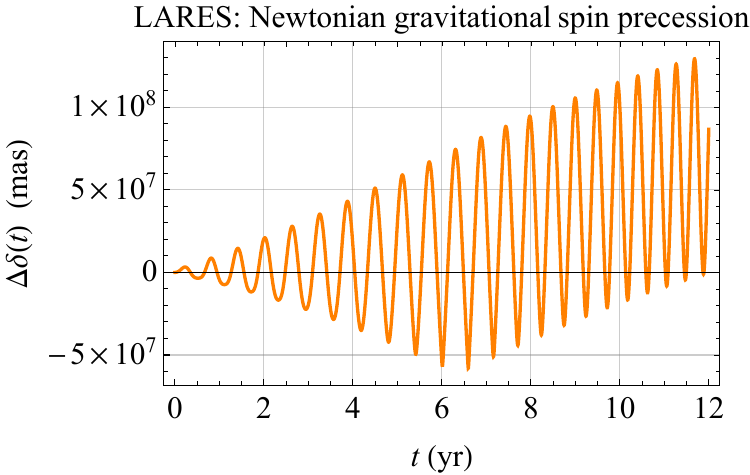}\\
\end{tabular}
\caption{
Numerically integrated time series, in mas, of the dS (upper row), PS (middle row) and Newtonian (lower row) RA and decl. shifts $\Delta\alpha\ton{t}$ and $\Delta\delta\ton{t}$ of the spin axis of LARES over 12 yr. The initial values of the spin and orbital parameters were retrieved from Table\,\ref{Tab:1}.
For the satellite's spin period $P_\mathrm{s}$, entering the Newtonian shifts, a linearly varying model \cite{2018PhRvD..98d4034V} according to the values of Table\,\ref{Tab:1} for $\dot P_\mathrm{s}$ was adopted. }\label{Fig:3}
\end{figure}

Figures \ref{Fig:1} to \ref{Fig:3} show the numerically integrated  time series of the dS, PS and Newtonian shifts of the RA and decl. of the rotational axes of LAGEOS, LAGEOS 2 and LARES.
For each satellite, they were obtained by numerically integrating any of the spin rate equations out of \rfr{Sdot_dS}, \rfr{Sdot_PS} and \rfr{Sdot_Qs} simultaneously with \rfrs{dIdt}{dOdt}.
The time span of the integrations are 49 yr for LAGEOS, 33 yr for LAGEOS 2 and 12 yr for LARES, respectively.

As far as the pN effects are concerned, it turns out that the RA grows steadily with oscillations superimposed on the main trends, while the decl. exhibits oscillatory patterns. The peak-to-peak amplitudes of the decl. signals amounts to about a thousand (dS) and five (PS) milliarcseconds (mas), respectively. Instead, the growing RA signatures amount to a few tens of thousands (dS) and a few hundreds (PS) of mas, respectively, over the time spans of the integrations.
Instead, the Newtonian shifts, obtained by including the slowdown of the satellites' spinning periods as per \cite{2018PhRvD..98d4034V}, are clearly oscillatory, especially the RA signals, with peak-to-peak nominal amplitudes as large as hundreds of millions of mas. \textcolor{black}{Such a peculiar feature turns out to be caused just by the lengthening of $P_\mathrm{s}$ modelled as an increasing linear trend according to Table\,\ref{Tab:1} and \cite{2018PhRvD..98d4034V}. Should it be neglected by keeping the rotational periods fixed, cleaner and regular trends would be obtained. It has been decided not to show them since they would be unrealistic, at least in the case of LAGEOS, LAGEOS 2 and LARES.}

Actually, the spin axis of a LAGEOS-type satellite is displaced also by a number of further torques of non-gravitational origin \cite{1991JGR....96.2431B,1996JGR...10117861F,1996GeoRL..23.3079V,2004JGRB..109.6403A,2018PhRvD..98d4034V}. One of them is due to the interaction of the Earth's magnetic field ${\bds B}_\oplus$ with the magnetic dipole moment ${\boldsymbol{\mu}}_\mathrm{s}$ acquired by the satellite during its diurnal rotation and orbital revolution through ${\bds B}_\oplus$ itself. Indeed,  being a metallic object, the spacecraft is a conductor; thus, its motions in the external magnetic field of our planet induce parasitic eddy electrical currents within its body which are just the source of ${\boldsymbol{\mu}}_\mathrm{s}$. The possible non coincidence between the satellite's Centre of Mass (CoM) and its geometrical center causes a further torque. Finally, another torque arises from the difference in the reflectivity of the two satellite's hemispheres.
A detailed analysis of their impact on the proposed measurement of the pN effects is outside the scopes of the present paper.

At present,  the spin axis orientation and/or rate of LAGEOS and its cousins are measured by exploiting the fact that either solar or laser radiation hitting the satellite is bounced back by its Corner Cube Retroreflectors (CCRs) which are oriented in rows. The number of spikes can be assessed by performing a frequency analysis  so that the spin rate is estimated. The orientation of the satellite's rotation axis can be inferred by combining the returns of the impinging electromagnetic waves with the geometry of the CCRs and that of the light source, satellite and observatory  \cite{1997dana.coll..341C,2001GeoRL..28.2113B,2004ITGRS..42..202O}.  Modern SLR systems characterized by high-repetition rate typically greater than 1 kHz and short pulse duration of the order of 10 ps reach a sub-degree accuracy level of about $0.1^\circ$. It may be difficult to improve this limit since the optical \virg{range} does not translate  well to the orientation. Instead, collecting several solar glints during a single pass can lead to a very accurate determination of the satellite's spin axis orientation, likely one or two orders of magnitude better than SLR. Such an approach is successfully exploited with the fast spinning Ajisai satellite \cite{1987ITGRS..25..526S}; in the case of the LAGEOS family, one may use ground based photon counters to try to detect solar reflections off the satellite's structure. In the future, continuous wave lasers and interference phenomena may allow for improvements in the accuracy of the laser-based spin determination of slowly rotating SLR satellites. In general, the faster the satellite rotates, the more stable its spin and the more accurately its orientation can be measured. Furthermore, a rapid rotation rate also allows  to reduce the impact of the Newtonian gravitational torque, as per \rfr{AJ2}.

\begin{figure}[ht!]
\centering
\begin{tabular}{cC}
\includegraphics[width = 8.5 cm]{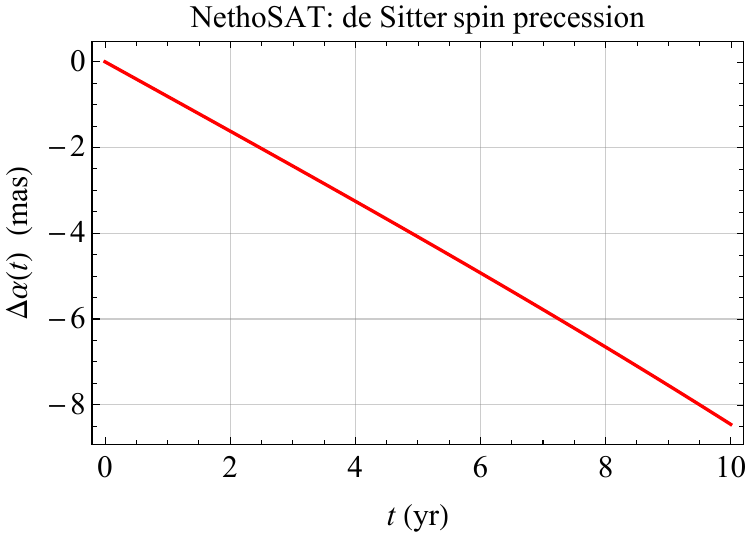} & \includegraphics[width = 8.5 cm]{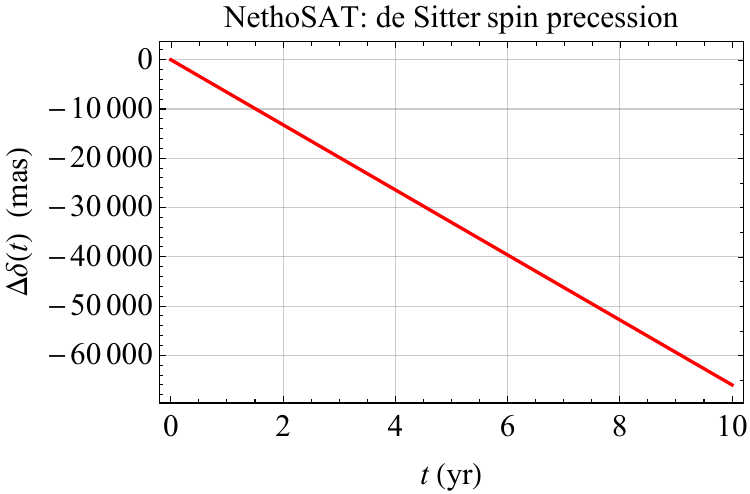}\\
\includegraphics[width = 8.5 cm]{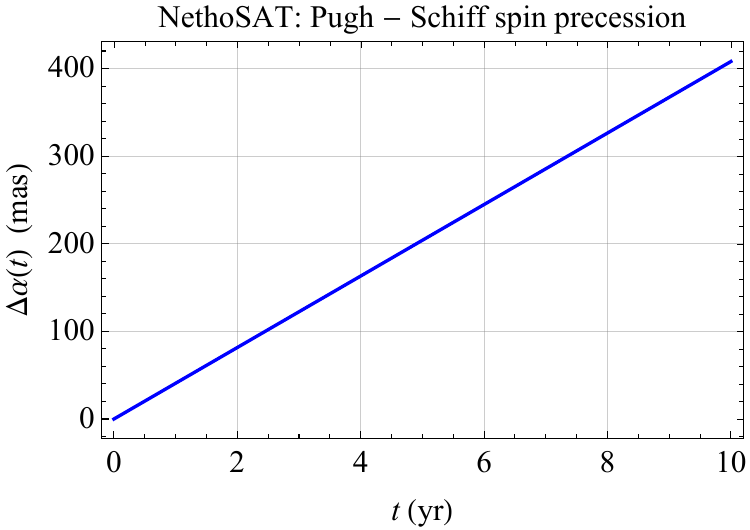} & \includegraphics[width = 8.5 cm]{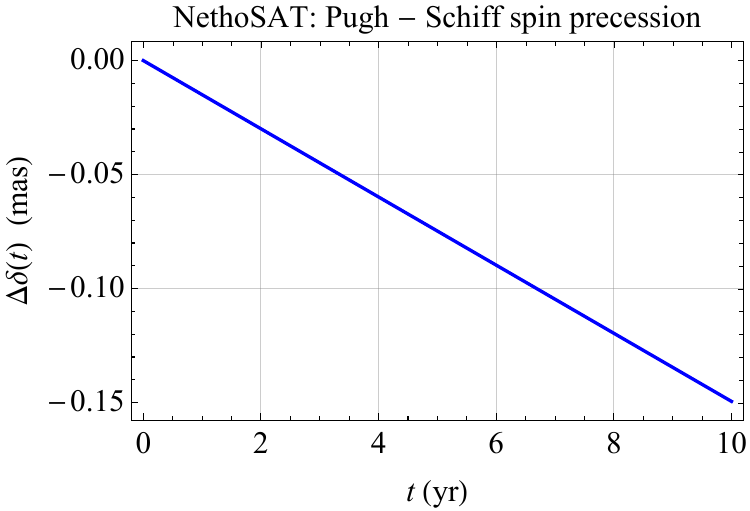}\\
\end{tabular}
\caption{
Numerically integrated time series, in mas, of the dS (upper row) and PS (lower row)  RA and decl. shifts $\Delta\alpha\ton{t}$ and $\Delta\delta\ton{t}$ of the spin axis of a hypothetical NethoSAT over 10 yr. The initial values of the spin and orbital parameters of GP-B, retrieved from Table\,\ref{Tab:1}, were adopted.
}\label{Fig:4}
\end{figure}

Figure\,\ref{Fig:4} displays the numerically integrated pN shifts of the spin axis of a hypothetical new satellite, dubbed NethoSAT (NS), with the same spin and orbital configurations of GP-B.
It can be noted that while the largest pN signature affecting the RA is a gravitomagnetic PS linear trend with a slope of 40 mas/yr, the decl. exhibits a huge graviteoectric dS precession of $-7000$ mas /yr. Should NS be a passive geodetic SLR satellite, a careful manufacturing and ground testing of it may, in principle, reduce the competing classical effects to an acceptable level.

\section{The pulsar-supermassive black hole scenario}\lb{Sec:3}

Here, the primary will be an extremely massive rotating black hole, and the role of nethotron will be played by a pulsar orbiting it.
\textcolor{black}{
As far as the effect of the environment (radiation, dust or even dark energy) in the timing of a pulsar orbiting Sgr A$^\ast$, see, e.g., \cite{2023PhRvD.108l4027C}.
}
Some relevant physical properties of black holes and neutron stars are reviewed below.

According to the no-hair theorems \cite{1967PhRv..164.1776I,1971PhRvL..26..331C,1975PhRvL..34..905R},
the mass and the spin moments  $\mathcal{M}_\bullet^\ell$  and  $\mathcal{J}_\bullet^\ell$  of degree $\ell$ \cite{1970JMP....11.2580G,1974JMP....15...46H} of a spinning black hole \citep{1970Natur.226...64B}, whose external spacetime is described by the Kerr metric \citep{1963PhRvL..11..237K,2015CQGra..32l4006T},  are connected by the relation
\eqi
\mathcal{M}_\bullet^\ell + j\,\mathcal{J}_\bullet^\ell= M_\bullet \ton{j\,\rp{J_\bullet}{c M_\bullet}}^\ell,\lb{nohair}
\eqf
where $j:=\sqrt{-1}$ is the imaginary unit. From \rfr{nohair}, it turns out that the odd mass moments and even spin moments
are identically zero. In particular, the mass moment $\mathcal{M}_\bullet^0$ of degree $\ell=0$  is the hole's mass $M_\bullet$, while its spin dipole moment $\mathcal{J}_\bullet$ of degree $\ell=1$ is proportional to its spin angular momentum $J_\bullet$. For  a Kerr BH, it is \cite{1986bhwd.book.....S}
\begin{align}
J_\bullet & = \chi_\bullet \rp{M_\bullet^2 G}{c},\,\left|\chi_\bullet\right|\leq 1,\acap
Q_2^\bullet & = -\rp{J_\bullet^2}{c^2 M_\bullet},
\end{align}
where $Q_2^\bullet$ is the mass moment $\mathcal{M}_\bullet^2$ of degree $\ell=2$.
If $\left|\chi_\bullet\right|>1$, a naked singularity \cite{1973CMaPh..34..135Y,1991PhRvL..66..994S} without a horizon would form, so that causality violations because of closed timelike curves may occur. The cosmic censorship conjecture \cite{1999JApA...20..233P,2002GReGr..34.1141P} has not yet been proven; nonetheless, it states that naked singularities may not be formed via the gravitational collapse of a material body.
In the following, the SMBH at the GC with $M_\bullet = 4.1\times 10^6\,M_\odot,\,\chi_\bullet =0.5$ \cite{2022ApJ...933...49P} will be considered as primary. However, it should be noted that a large uncertainty still plagues the value of $\chi_\bullet$ for Sgr A$^\ast$ which may be as large as $\sim 0.9$ according to some recent results \cite{2023Astro...2..141D,2024ARep...68..233A,2024MNRAS.527..428D}.

The dimensionless quadrupole mass moment $J_2^\star$ of a rotating neutron star can be expressed in terms of its dimensional counterpart $Q_2^\star$ having dimensions of a mass times a length squared as
\eqi
J_2^\star = -\rp{Q^\star_2}{M_\star R_\star^2},\lb{J2psr}
\eqf
where \cite{1999ApJ...512..282L}
\eqi
Q^\star_2 = \xi_\star\rp{G^2 M_\star^3}{c^4}.\lb{Q2psr}
\eqf
The size of the parameter $\xi_\star$ ranges within
\eqi
0.074\lesssim |\xi_\star| \lesssim 3.507\lb{xipsr}
\eqf
for a variety of equations of state (EOS) and $M_\star = 1.4\,M_\odot$ \cite{1999ApJ...512..282L}.
In the following, a millisecond pulsar with  $M_\star = 1.4\,M_\odot,\,R_\star=10\,\mathrm{km},\,\xi_\star=3.507,\,P_\star = 5\,\mathrm{ms}$ will be  considered as  nethotron orbiting the SMBH in Sgr A$^\ast$.

In such a scenario, yet to be discovered, predicting the pN spin variations is challenging because all the parameters characterizing the spin and orbital configurations of such a system are virtually unconstrained. It may be useful to recall them: they are the two polar angles $\zeta_\bullet, i_\bullet$ of the SMBH's spin axis, the RA $\alpha$ and DEC $\delta$ of the pulsar' spin axis, and the inclination $I$ and the longitude of the ascending node $\Omega$ characterizing the pulsar's orbital plane.

As far as the holes's spin axis ${\bds{\hat{J}}}_\bullet$ is concerned, in the following, it will be parameterized as
\begin{align}
\hat{J}^\bullet_x & = \cos\zeta_\bullet\sin i_\bullet, \acap
\hat{J}^\bullet_y & = \sin\zeta_\bullet\sin i_\bullet, \acap
\hat{J}^\bullet_z & = \cos i_\bullet,
\end{align}
where $i_\bullet$ is the tilt of $\boldsymbol{\hat{J}}_\bullet$ with respect to the line of sight, usually assumed as reference $z$ axis.
It turns out that $i_\bullet,\,\zeta_\bullet$ are still basically unconstrained.
Performed attempts to somehow constrain ${\boldsymbol{\hat{J}}}_\bullet$ of Sgr A$^\ast$\index{subject}{Sgr A$^\ast$} with different nondynamical approaches can be found in \cite{2007ApJ...662L..15F,2007A&A...473..707M,2009ApJ...697...45B,2011ApJ...735..110B,2012ApJ...755..133S,2016MNRAS.458.3614J,2016ApJ...827..114Y}, and references therein. On the one hand, it appears that $\boldsymbol{\hat{J}}_\bullet$ would not be aligned with the line of sight.
Indeed, according to, e.g., \cite{2007A&A...473..707M}, who used polarimetric observations of the near--infrared emission
of Sgr A$^\ast$, it is $i_\bullet\simeq 55^\circ$. \cite{2007ApJ...662L..15F} obtained $i_\bullet\simeq 77^\circ$ on the basis of
their fit of a simulated Rossby wave--induced spiral pattern in the hole's accretion disk to the X--ray lightcurve detected with
the mission X-ray Multi-Mirror-Newton (XMM--Newton). The authors of \cite{2012ApJ...755..133S} provided the range $42^\circ\lesssim i_\bullet \lesssim 75^\circ$ by comparing polarized submillimetre infrared observations with spectra computed using three-dimensional general relativistic  magnetohydrodynamical simulations.
Methods based on gravitational lensing for determining ${\bds{\hat{J}}}_\bullet$ independently of orbital dynamics were outlined, e.g.,
in \cite{2017PTEP.2017e3E02S}. The authors of \cite{2004ApJ...611..996T} investigated the possibility of measuring, among
other things, $i_\bullet$ from the shape and position of the hole's shadow under certain assumptions. On the other hand, the first observations by the Event Horizon Telescope (EHT) \cite{2022ApJ...930L..12E} disfavor, among other things, scenarios where the BH is viewed at high inclination  $\ton{i_\bullet >50^\circ}$.

It turns out that both the dS and the PS spin rates can vanish for certain values of the parameter space: the are
\begin{align}
\left|\dert{\boldsymbol{\hat{S}}}{t}\right|_\mathrm{dS}^\mathrm{min} &= 0,\,\alpha^\mathrm{min} = 211.2^\circ,\,\delta^\mathrm{min} = 37.3^\circ,\,I^\mathrm{min} = 127.3^\circ,\,\Omega^\mathrm{min} = 121.2^\circ, \acap
\left|\dert{\boldsymbol{\hat{S}}}{t}\right|^\mathrm{min}_\mathrm{PS} &= 0,\,\alpha^\mathrm{min} = 51.3^\circ,\,\delta^\mathrm{min} = -65.5^\circ,\,\zeta_\bullet^\mathrm{min} = 155^\circ,\,i_\bullet^\mathrm{min} = 116.5^\circ,\,I^\mathrm{min} = 27.3^\circ,\,\Omega^\mathrm{min} = 24.2^\circ.
\end{align}
The maximum values for the pN spin rates are obtained for
\begin{align}
\rp{1}{{\mathcal{A}}_\mathrm{dS}}\left|\dert{\boldsymbol{\hat{S}}}{t}\right|_\mathrm{dS}^\mathrm{max} \lb{dSmax}&= 1,\,\alpha^\mathrm{max} = 324.2^\circ,\,\delta^\mathrm{max} = 68^\circ,\,I^\mathrm{max} = 95.5^\circ,\,\Omega^\mathrm{max} = 338.2^\circ,\acap
\rp{1}{{\mathcal{A}}_\mathrm{PS}}\left|\dert{\boldsymbol{\hat{S}}}{t}\right|_\mathrm{PS}^\mathrm{max} \lb{LTmax} &= 1,\,\alpha^\mathrm{max} = 279.3^\circ,\,\delta^\mathrm{max} = 12.1^\circ,\,\zeta_\bullet^\mathrm{max} = 184.2^\circ,\,i_\bullet^\mathrm{max} = 67.2^\circ,\,I^\mathrm{max} = 67.2^\circ,\,\Omega^\mathrm{max} = 274.2^\circ.
\end{align}

\begin{figure}[ht!]
\centering
\begin{tabular}{cC}
\includegraphics[width = 8.5 cm]{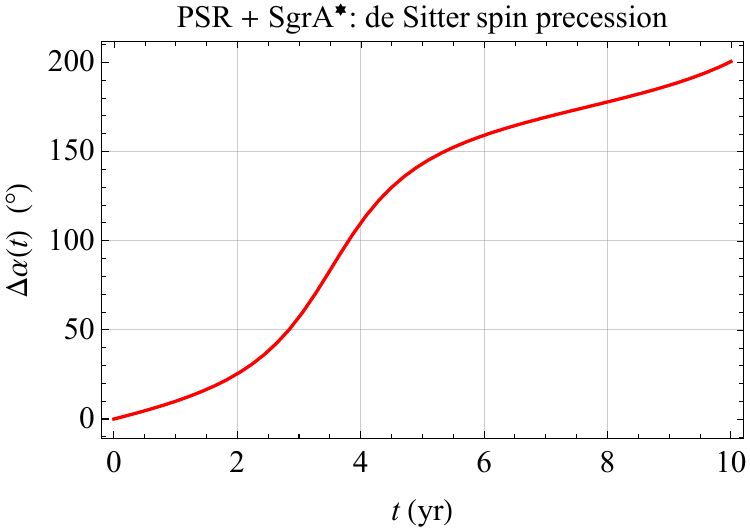} & \includegraphics[width = 8.5 cm]{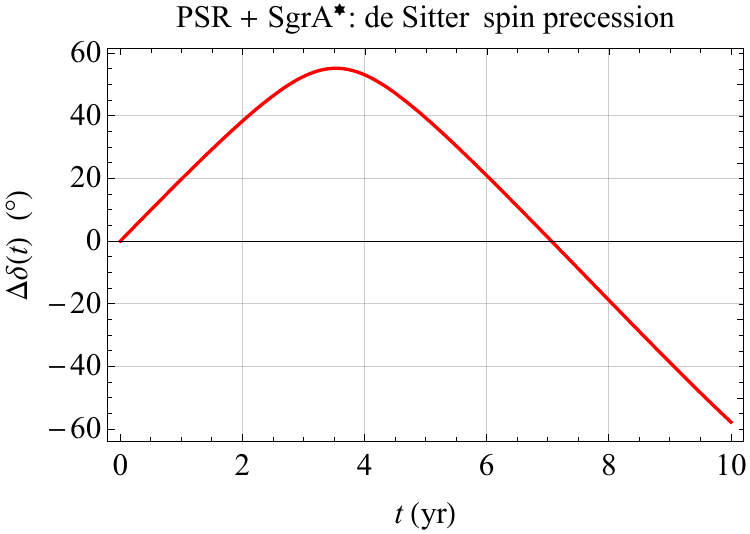}\\
\includegraphics[width = 8.5 cm]{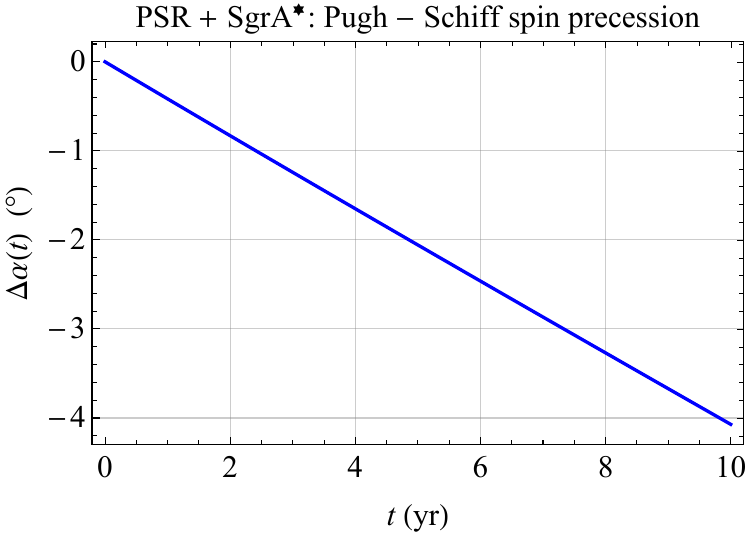} & \includegraphics[width = 8.5 cm]{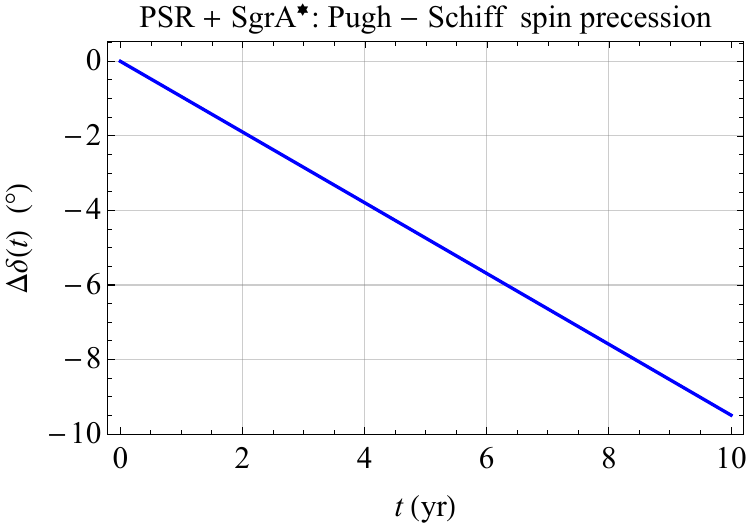}\\
\includegraphics[width = 8.5 cm]{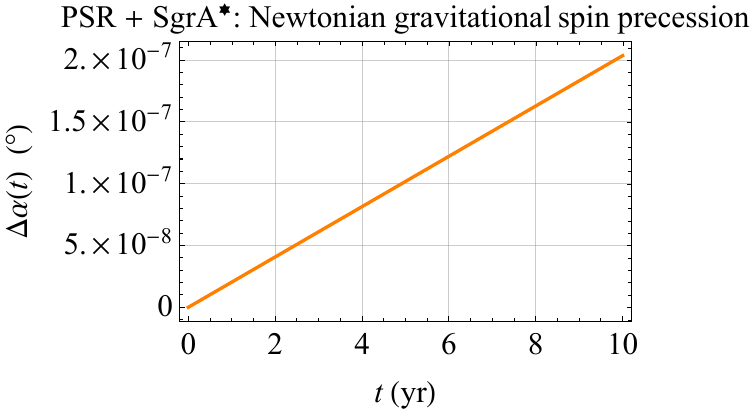} & \includegraphics[width = 8.5 cm]{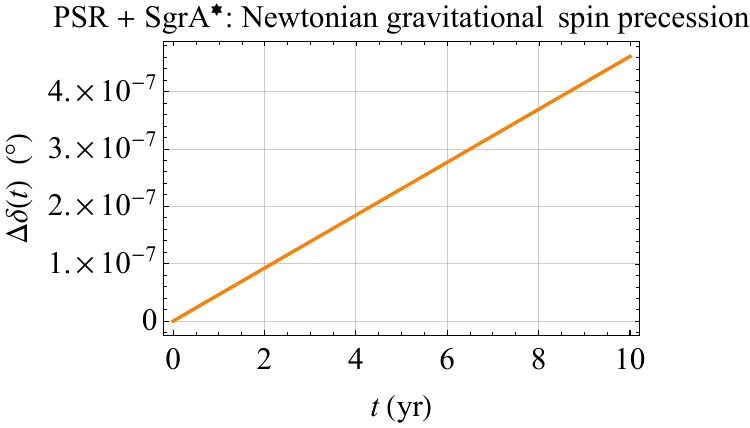}\\
\end{tabular}
\caption{
Numerically integrated time series, in $^\circ$, of the dS (upper row), PS (middle row) and Newtonian (lower row) RA and decl. shifts $\Delta\alpha\ton{t}$ and $\Delta\delta\ton{t}$ of the spin axis of a millisecond pulsar in a $0.5$ yr orbit around the SMBH in Sgr A$^\ast$ over 10 yr. While the eccentricity was chosen in order to have $r_\mathrm{min}= 12.4\,r_\mathrm{Sch}$, the other relevant spin and orbital parameters were retrieved from \rfr{LTmax}. For the pulsar's quadrupole parameter, the value $\xi_\star = 3.507$ was adopted in \rfr{Q2psr}.
}\label{Fig:5}
\end{figure}

Figure\,\ref{Fig:5} displays the numerically integrated pN signatures over 10 yr of the RA and decl. of the spin of a hypothetical millisecond pulsar orbiting the SMBH in Sgr A$^\ast$ along a highly eccentric orbit completed in $0.5$ yr which brings the neutron star as close as $\simeq 12$ Schwarzschild radii\footnote{The Schwarzschild radius is defined as $r_\mathrm{Sch}: = 2\upmu_\bullet/c^2$.} to the hole. Given that the system's parameter space is unconstrained, the values of, say,  \rfr{LTmax} were adopted for the spin and orbital configurations. As far as the pulsar's quadrupole mass moment is concerned, whose signature is depicted in Figure\,\ref{Fig:5} as well, it was calculated according to \rfr{Q2psr} with $\xi_\star =3.507$.
The size of the pN signatures is remarkable, amounting to tens or hundreds of degrees. Remarkably, the effect of the gravitational torque due to the pulsar's own oblateness turns out to be completely negligible already at the Newtonian order\footnote{In fact, it is of the order of $\mathcal{O}\ton{c^{-4}}$, as per \rfr{Q2psr}.}. This confirms that the nethotron approximation is also valid for such a compact object in a strong external gravitational field.

If, on the one hand, the non-gravitational torques induced by the possible CoM offset and the asymmetric reflectivity affecting the LAGEOS satellites are absent here, on the other hand the magnetic torque should be a major competing effect also in this scenario. Indeed, it is well known that pulsars posses extremely huge magnetic moments giving rise to strengths of the magnetic field at their surfaces as large as  $B_\star\simeq 10^8 - 10^{14}$ G \cite{2023Univ....9..334K}. Furthermore, around the SMBH at the GC there is a strong magnetic field \cite{2013Natur.501..391E,2024ApJ...964L..25E}.
In fact, even if the external magnetic field were absent, other electromagnetic torques would affect the pulsar's spin axis as well because of the misalignment between the latter and the magnetic dipole moment ${\bds \mu}_\star$. Indeed, a spherical body endowed with a magnetic dipole field rotating in vacuum experiences two torques acting on it \cite{2015MNRAS.451..695Z}. The first one arises from the fact that a misaligned spinning magnetic dipole moment emits electromagnetic waves which carry away angular momentum \cite{2015MNRAS.451..695Z}. The other one is due to the inertia of ${\bds \mu}_\star$ \cite{1970ApJ...159L..81D}. Both are proportional just to ${\bds\mu}_\star\bds\times\bds{\hat{S}}_\star$ \cite{2015MNRAS.451..695Z}.
An evaluation of such  potentially relevant competing effects is outside the scopes of the present work.
\section{The double pulsar}\lb{Sec:4}
Soon after its discovery in 2003 \cite{2003Natur.426..531B,2004Sci...303.1153L}, the double pulsar PSR J0737-3039A/B \cite{2008ARA&A..46..541K} became
a celestial laboratory of primary importance to test gravitational physics in the strong field regime, primarily through its orbital dynamics and the effects on travelling electromagnetic waves \cite{2006Sci...314...97K,2009CQGra..26g3001K}.
As far as the spin effects are concerned, if, on the one hand, it allowed to measure the generalization of the gravitoelectric dS-or spin-orbit (SO) in pulsar's literature-precession to $\simeq 13$ per cent  \cite{2008Sci...321..104B} and $\simeq 6$ per cent accuracy \cite{2024A&A...682A..26L}, on the other hand, there are no reports of attempts made or proposed to measure the gravitomagnetic-or spin-spin (SS) in pulsar's literature-shift.

The authors of \cite{2008Sci...321..104B,2024A&A...682A..26L} determined the magnitude of the angular velocity ${\bds\Omega}^\mathrm{B}$ of the spin precession of the B component of the double pulsar to an accuracy of
\eqi
\sigma_{\Omega^\mathrm{B}}\simeq 0.6-0.3^\circ/\mathrm{yr}.\lb{errsB}
\eqf
Here, the predicted magnitude of the angular velocity of the gravitomagnetic SS precession of the B's spin will be calculated and compared with \rfr{errsB}.

In the case of a binary system composed of two bodies with comparable masses and angular momenta, the gravitomagnetic precession of the spin of one of them induced by the spin of the other one can be expressed by \rfrs{Sdot_PS}{OPS} provided that the orbital parameters and vectors are referred to the binary's relative motion \cite{1975PhRvD..12..329B}. In the particular case of the double pulsar,  $\bds S$ entering \rfr{Sdot_PS} represents the precessing angular momentum of B, while $\bds J$ entering \rfr{OPS} is meant as the angular momentum of A. It is so because $S_\mathrm{B}/S_\mathrm{A}\simeq 0.008$, so that the gravitomagnetic  precession of the spin of A due to the spin of B is negligible. Indeed, while the spin period of A is $P_\mathrm{A} = 22$ ms, the one of B amounts to $P_\mathrm{B} = 2.77$ s \cite{2006Sci...314...97K}. The moments of inertia $\mathcal{I}$ of both neutron star were assumed equal to $\mathcal{I}\simeq 1.6\times 10^{38}\,\mathrm{kg}\,\mathrm{m^2}$ (see \cite{2021PhRvL.126r1101S}, and references therein). By parameterizing the A's spin axis ${\bds{\hat{J}}}_\mathrm{A}$ in terms of its colatitude $\lambda_\mathrm{A}$ and azimuth angle $\psi_\mathrm{A} $ in the plane of the sky with respect to the reference $x$ direction in it, from \rfr{OPS} it turns out
\eqi
\Omega^\mathrm{B}_\mathrm{SS}\simeq 0.00008^\circ/\mathrm{yr},\lb{OMBO}
\eqf
which is about 4 orders of magnitude smaller than \rfr{errsB}. In order to calculate \rfr{OMBO}, the range of values obtained in \cite{2021MNRAS.507..421I} for the A's spin direction was adopted.
\section{Summary and conclusions}\lb{Sec:5}
Measuring the temporal evolution of the orientation of the spin axes of natural and/or artificial nethotrons like pulsars orbiting the supermassive black hole in the Galactic Centre and Earth's laser-ranged satellites  may represent, in principle, a relatively low-cost approach to strengthen the empirical basis supporting the post-Newtonian contributions to such kind of phenomena. Indeed, the sole experiment which measured so far the gravitomagnetic Pugh-Schiff spin precession is GP-B which will likely never be repeated due to its cost and complexity.

Thus, it seems worth of investigating the possibilities offered by the existing passive geodetic satellites of the LAGEOS type, whose spin axis orientation has been the subject of long and intense observation campaigns  performed with different techniques for many years. Measuring their de Sitter and Pugh-Schiff spin shifts seems, at present, a daunting task. Indeed, the current  accuracy in measuring the evolution of the orientation of their rotational axes over the years, being at the $\simeq 0.1^\circ$ level in best cases, is low with respect to the expected sizes of the post-Newtonian signals, of the order of a few tens of thousands or hundreds of milliarcseconds. Furthermore, the Newtonian gravitational torque due to the satellites' own oblateness
is a major source of systematic bias since its nominal effects are several orders of magnitude larger than the relativistic ones. Last but not least, also non-gravitational torques like, e.g., that arising from the interaction of the Earth's magnetic field with the acquired magnetic dipole moments of the satellites, enter the error budget. The situation may become more favorable in the future should one or more carefully manufactured satellites, provisionally dubbed NethoSAT(s), be launched and new will observational techniques be devised.

The size of the relativistic spin precessions of natural nethotrons like pulsars orbiting the supermassive black hole in Sgr A$^\ast$ at the Galactic Centre, long sought because of their high potential as effective probes of the hole's spacetime, may be as large as some tens or hundreds of degrees for orbital periods as short as $0.5$ yr. In this scenario, major competing effects would be those driven by various electromagnetic torques. Instead, the Newtonian gravitational spin shift attributable to the pulsar's own quadrupole mass moment would not be of concern. It should be recalled that only the orbital precessions of such hypothesized pulsars have been considered so far as means to probe the spacetime features of Sgr A$^\ast$. Measuring also their spin precessions would be an additional resource to aim at such a goal.

As far as the double pulsar is concerned, the gravitomagnetic precession of the spin of its B component due to the A's angular momentum is expected to be about 4 orders of magnitude smaller than the current level of accuracy in measuring it.

\section*{Data availability}
No new data were generated or analysed in support of this research.
\section*{Conflict of interest statement}
I declare no conflicts of interest.
\section*{Funding}
This research received no external funding.
\section*{Acknowledgements}
I am grateful to D. Kucharski for having provided me with important information on the measurement techniques of the spin axes of SLR satellites.

\bibliography{Megabib}{}
\end{document}